\documentclass[10pt,twocolumn,preprintnumbers,amsmath,amssymb,nofootinbib,superscriptaddress]{revtex4-1}

\usepackage{graphicx}
\usepackage{dcolumn}
\usepackage{amssymb,amsmath,bm}
\usepackage{color}
\usepackage[colorlinks,linkcolor=red,citecolor=blue,urlcolor=blue ]{hyperref}
\usepackage{multirow}
\usepackage{enumitem}
\usepackage{mathptmx} 
\usepackage[utf8]{inputenc}
\usepackage{tensor}
\usepackage{amssymb} 
\usepackage[normalem]{ulem}
\usepackage{lettrine}
\usepackage{empheq}
\usepackage{comment}


\begin{document}

  \title{The impact of initial conditions on quasi-normal modes}
   \author{Ameya Chavda}
  \email{ameya.chavda@columbia.edu}
  \affiliation{Center for Theoretical Physics, Departments of Physics and Astronomy, Columbia University, New York, NY 10027, USA}
   \author{Macarena Lagos}
  \email{macarena.lagos.u@unab.cl}
  \affiliation{Instituto de Astrof\'isica, Departamento de F\'isica y Astronom\'ia, Universidad Andr\'es Bello, Santiago, Chile}
   \author{Lam Hui}
  \email{lh399@columbia.edu}
   \affiliation{Center for Theoretical Physics, Departments of Physics and Astronomy, Columbia University, New York, NY 10027, USA}

\begin{abstract}
This study investigates the influence of initial conditions on the evolution and properties of linear quasi-normal modes (QNMs). Using a toy model in which the quasi-normal mode can be unambiguously identified, we highlight an aspect of QNMs that is long known yet often ignored: the amplitude of a QNM (after factoring out the corresponding exponential with a complex frequency) is not constant but instead varies with time. We stress that this is true even within the regime of validity of linear perturbation theory. The precise time variation depends on the initial conditions. In particular, it is possible to find initial conditions for which the QNM fails to materialize; it is also possible to find those for which the QNM amplitude grows indefinitely. Focusing on cases where the QNM amplitude does
stabilize at late times, we explore how the timescale for amplitude stabilization depends on the shape and location of the initial perturbation profile. Our findings underscore the need for care in fitting linear QNMs to ringdown data. They also suggest recent computations of quadratic QNMs, sourced purely by {\it stabilized} linear QNMs, do not fully capture what determines the amplitude of the quadratic QNMs, even at late times. Our results motivate a detailed investigation of the initial perturbations generated in the aftermath of a binary merger.

\end{abstract}

  \maketitle

\section{Introduction}\label{sec:intro}

The detection of gravitational waves from binary mergers has opened the window to study the astrophysics and fundamental physics governing such events \cite{LIGOScientific:2016vbw}
While numerical general relativistic simulations have become the indispensable tool for interpreting the gravitational wave form, perturbation theory---around the background of the expected merger product, that is, a remnant black hole (BH)---remains a powerful tool for understanding the ring-down signal \cite{PhysRev.108.1063,PhysRevD.2.2141,Teukolsky:1972my}. 
This is especially true if one wishes to test general relativity (GR) in a model independent way \cite{Dreyer:2003bv,Cardoso:2016ryw,Franciolini:2018uyq,Isi:2019aib,Hui:2021cpm}
It's worth stressing that ``testing GR'' is an umbrella term, which includes
testing the fundamental law of gravity, as well as searching for modifications of the black hole environment, due to the presence of dark matter or a superradiance cloud for instance.


Fitting for quasinormal modes (QNMs), each characterized by a complex frequency, is the standard procedure for analyzing ring-down data \cite{Kokkotas:1999bd,Ferrari:2007dd,Berti:2009kk}.
As the data quality improves, it is important we improve the understanding of BH perturbation theory: its regime of validity and what exactly it predicts.
Regarding the former, there has been some discussion of the appropriate start-time for analyzing the ring-down data\cite{Bhagwat_2018,Giesler:2019uxc,Baibhav:2017jhs,Baibhav:2023clw}.
Related to this discussion is the realization that 
second-order QNMs, a prediction of second-order perturbation theory \cite{Gleiser:1995gx, Nakano:2007cj, Ioka:2007ak, Bhagwat:2017tkm,Okuzumi:2008ej,  Okounkova:2020vwu, Sberna:2021eui, Lagos:2022otp, Ma:2024qcv,Bourg:2024jme, Bucciotti:2024zyp,Bucciotti:2024jrv}, can be unambiguously identified in numerical GR simulations \cite{Ma:2022wpv, Cheung:2022rbm, Mitman:2022qdl, Cheung:2023vki, Yi:2024elj}.

In this paper, we focus exclusively on first order perturbation theory and what it predicts. It might surprise the reader that its precise predictions are not {\it well} known. After all, it's an old subject. Let's recall the setup. The linear perturbation around a BH background, signified here by $\Psi$, obeys a differential equation of the form:
\begin{eqnarray}
\label{Vgeneral}
(-\partial_t^2 + \partial_x^2 - V(x)) \Psi(t, x) = 0 \, .
\end{eqnarray}
Here, $x$ is the tortoise coordinate with the horizon at $x=-\infty$. An implicit separation of variables has been assumed, in which case $V(x)$ would depend on the angular momentum numbers
$\ell, m$.
\footnote{\label{footnote1} In other words, $\Psi(t, x)$ implicitly depends on $\ell, m$. To obtain the full 4D space-time dependence, one superimposes: $\sum_{\ell, m} \Psi(t, x) Y_{\ell m} (\theta, \phi)$
  where $Y_{\ell m}$ represents the spherical harmonics for Schwarzschild BH (or the spheroidal harmonics for Kerr BH).
  }
For a Schwarzchild black hole, the $m$ dependence would be absent, and
$V$ would be the Regge-Wheeler or Zerilli potential, with $t$ being the Schwarzschild time
\cite{PhysRev.108.1063,PhysRevD.2.2141}.

The retarded Green function $G(t, x |\bar t, \bar x)$ can be used to evolve $\Psi$, as follows
(see e.g. \cite{1962clel.book.....J,Szpak:2004sf,Hui:2019aox}):
\begin{align}
    \Psi (t,x) =\int d\bar{x} \left[ \partial_{\bar{t}}G
      |_{\bar{t}=0}\Psi_{0}(\bar{x})- G |_{\bar{t}=0}\dot{\Psi}_{0}(\bar{x})\right] \label{Phi1_Gral}  .
\end{align}
Here, $\bar t=0$ represents the time at which the initial perturbation
$\Psi_0$ and its time derivative $\dot\Psi_0$ are given.
\footnote{Here, $\Psi_0$ and $\dot\Psi_0$ are also implicitly dependent on $\ell, m$.
The same comment as in footnote \ref{footnote1} applies to them.}
The retarded nature of $G$ means it vanishes unless $t > \bar t$.

\begin{figure} [h!] 
    \centering
    {\includegraphics[width=0.5\textwidth]{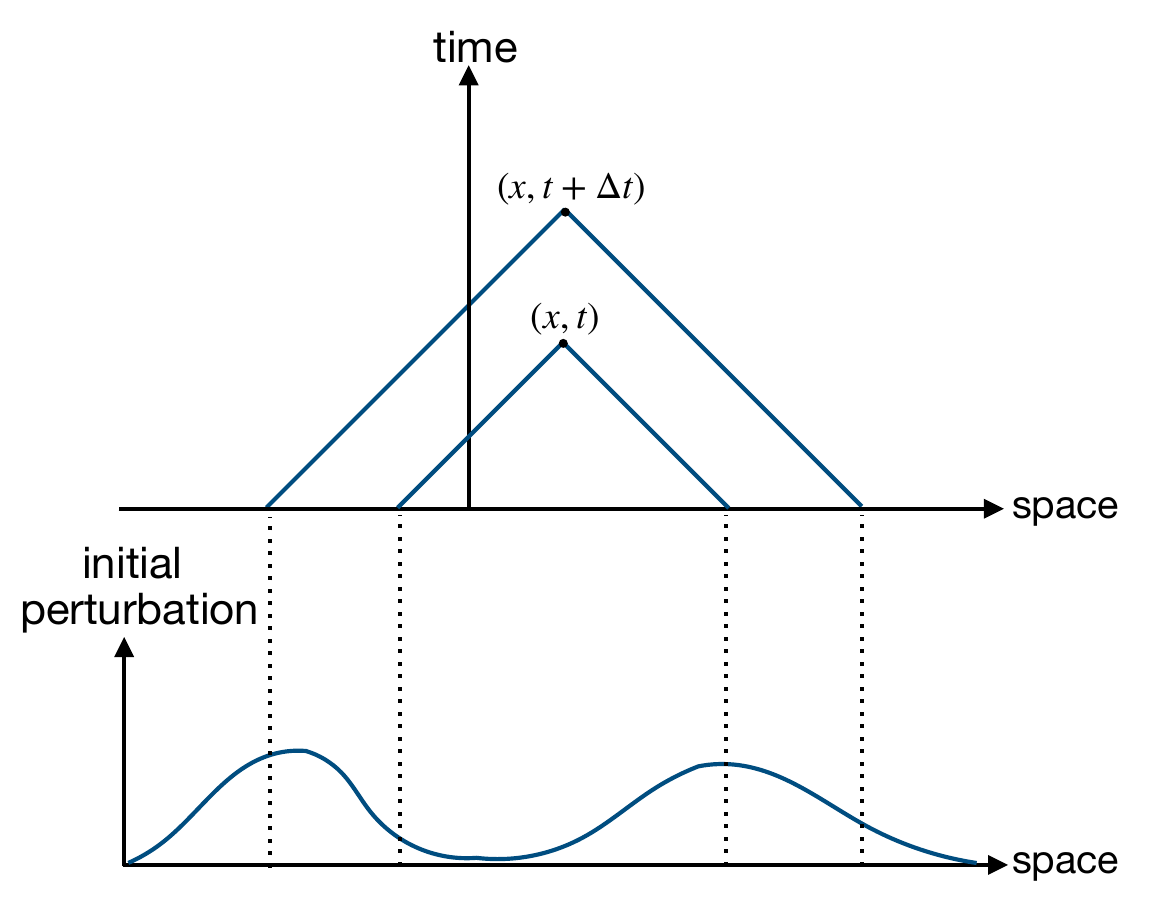}}
    \caption{A schematic diagram illustrating how the observed time evolution of (GW) perturbation depends on the initial condition. Imagine an observer at location $x$ observing
      the perturbation $\Psi$ at time $t$ and at time $t + \Delta t$. The top panel is a space-time diagram showing the respective backward lightcones from the two corresponding points in space-time. The Green function $G$ is expected to be non-vanishing only within each backward lightcone. The bottom panel is a schematic illustration (not to be taken literally) of the initial perturbation $\Psi_0$. It is evident that as time progresses, more of the initial perturbation becomes accessible within the backward lightcone, and thus affects the observed signal $\Psi$. See Fig. \ref{causality graph} on which portions of the initial perturbation are responsible for the QNMs.
    } 
    \label{QNMspacetime}
\end{figure}

It is well known that $G$ in Fourier space has poles at the quasinormal frequencies
\cite{PhysRevD.34.384, Andersson:1996cm,Szpak:2004sf,Lagos:2022otp},
which leads to the expectation that $\Psi(t, x)$ inherits the corresponding time dependence.
But it's important to keep in mind $G$ also has other structures (such as branch cut,
a pole or branch point at
vanishing frequency, and possible non-trivial contributions at large
complex frequencies). An important aspect of some of these structures is that they
enforce causality: that $G(t, x| \bar t, \bar x)$ vanishes if $t - \bar t > |x - \bar x|$.
\footnote{That $G$ is cut off outside the light-cone is important for
  understanding a puzzling aspect of QNMs. Consider an outgoing mode $e^{-i \omega (t - x)}$ far away from the BH. A QNM should have ${\rm Im\,}\omega < 0$ such that it decays with time $t$. This means the same mode would exponentially diverge at large $x$. Such unphysical behavior is absent from the Green function---even though it knows about the QNMs---by virtue of its causality structure.}
The situation is depicted schematically in Figure \ref{QNMspacetime}. Imagine an observer at a fixed location $x$.
As time $t$ progresses, a larger portion of the initial perturbations comes into view.
It's thus not surprising that the observed time-dependence is affected by the spatial
profile of the initial perturbations. Therefore, even though $\Psi(t, x)$ contains imprints of the
QNMs, their amplitude will in general vary with time, in a way that is sensitive to the spatial dependence of the initial perturbations.

It is important to stress this fact---that QNMs generally have time-dependent
amplitudes---has been discussed by a number of authors \cite{PhysRevD.34.384, Andersson:1996cm,Szpak:2004sf,Lagos:2022otp}.
Nonetheless, it is standard practice in the literature to fit ring-down data
(from both simulations and observations) with a series of QNMs, each with
a constant amplitude:
\begin{eqnarray}
\label{superimpose}
\sum_{\ell, m, n} A_{(\ell, m, n)} e^{-i\omega_{(\ell, m, n)} t} \, .
\end{eqnarray}
Here, $\omega_{(\ell, m, n)}$ corresponds to the QNM (complex) frequency
for angular momentum numbers $\ell, m$ and overtone $n$.
The constancy of each amplitude $A_{(\ell, m, n)}$ is often taken for granted.
\footnote{Indeed, sometimes it's even argued that non-constancy of the QNM amplitude,
when observed in fits to numerical simulations,
suggests a breakdown of linear perturbation theory.
As we will spell out, the non-constant nature
of each QNM amplitude is in fact a generic expectation,
even within linear theory.}
A natural question arises: to what extent is the observed constancy,
in fits to numerical simulations, an artifact of the fitting procedure?
Indeed, recent discussions have highlighted the effects of
various choices such as the fitting time, the number of overtones,
as well as the effects of noise and polynomial tails \cite{Giesler:2019uxc,Baibhav:2023clw}. 

We thus find it useful to return to the basics, and explore in detail a simple
example where the QNM is analytically known, freeing ourselves from
ambiguities having to do with the fitting procedure. We are interested in an example
where the Green function can be written down explicitly, and the corresponding
Cauchy problem (Eq.\ \ref{Phi1_Gral}) can be easily solved for a variety of initial conditions.
The delta function potential $V(x) = V_0 \delta (x)$, where $V_0$ is a constant, provides such an example.
It's admittedly a crude approximation to the actual potential (Regge-Wheeler for instance). For instance, the Regge-Wheeler potential has power-law tails, which the delta function obviously lacks.
Nonetheless, the fact that the QNM frequencies are well approximated by the WKB approximation \cite{1985ApJ...291L..33S}, wherein derivatives of the potential at its peak completely determine the QNM spectrum, suggests approximating the potential by a well-localized one (a delta function localized at the peak) might capture some of the salient qualitative features.

Our focus will be on the time-dependence of the QNM amplitude. {\it Are there initial conditions for which the QNM amplitude never asymptotes to a constant? For those that do, how long does one have to wait for the QNM amplitude to stabilize?} A recurring theme that emerges out of our investigation is that it would be very useful to study in detail the nature of the initial perturbations in the aftermath of a binary merger. In particular, as we will see, their location and width have implications for the amount of time it takes for the QNM amplitude to stabilize.

This paper is organized as follows.
In Section \ref{sec:delta} we summarize the toy model that will be used to study the linear evolution of gravitational perturbations, from which the QNMs emerge.
In Section \ref{sec:nQNMs}, we
study examples of initial conditions for which QNMs are somewhat elusive: either they fail to materialize, or if they do materialize, they would be difficult to identify and isolate. These examples are admittedly ad hoc, but they underscore the important point that the way QNMs emerge (or fail to emerge) out of the Cauchy problem depends crucially on the nature of the initial conditions. 
In Section \ref{sec:gaussiansection}, 
we study the evolution problem for initial perturbations that take a Gaussian form. We explore how the QNM amplitude, and its stabilization time, depend on the location and width of the Gaussian. We will see that the standard wait/stabilization time of $10 M$ is by no means guaranteed. In Section \ref{sec:stepIC}, we explore the same issues for initial perturbations that take the form of a hyperbolic tangent, for which perturbations do not vanish at the horizon, as suggested by binary merger simulations. We conclude in Section \ref{sec:conclusions} with a discussion of interesting questions to be explored in the future.

Throughout this paper, the speed of light $c$ and the gravitational constant $G$ are set to unity.
As we will explain below, the toy model we use has one dimensionful parameter $V_0$, which serves to normalize the potential, with $1/V_0$ defining a timescale.
Anticipating the QNM will turn out to have an imaginary frequency of $-V_0/2$ (i.e. its associated time dependence is $e^{-V_0 t /2}$), it's helpful to compare this against the $(2,2,0)$ QNM for a Schwarzschild BH, for which the imaginary part of the frequency is $-0.0889/M$ where $M$ is the BH mass. In our explorations below, we will thus sometimes translate our findings by equating $V_0$ to $0.18/M$, or conversely, equating a timescale of $1/V_0$ to $5.6 \, M$. The conversion is summarized in Table \ref{Params}.
\footnote{It is worth noting that the imaginary part of the QNM frequency does not appear to be hugely sensitive to the precise values of $\ell, m$, for Schwarszchild as well as moderately rotating Kerr BHs. There is a stronger dependence on the overtone number $n$. We will comment on the interpretation of our findings for $n \ne 0$ below.
}
\begin{table}[h!]
\centering
\begin{tabular}{ccc}
\hline
Conversion between $V_0$ and $M$ \\ \hline
$V_0 \approx 0.18 M^{-1}$\\
$\frac{1}{V_0} \approx 5.6 M$\\ \hline
\end{tabular}
\caption{This table indicates that when $V_0 t$ is depicted in a figure, a value of for instance $V_0 t = 1$ can be interpreted as $t \approx 5.6 M$. Similarly for $x$: $V_0 x = 1$ can be interpreted as $x \approx 5.6 M$.}
\label{Params}
\end{table}

One last comment on terminology. In this paper, we will be referring a lot to the QNM amplitude. By this, we mean something like the amplitude $A$ in Eq. (\ref{superimpose}), in which the expected exponential with a complex frequency is already factored out. The analog of this for our toy model would be $A$ in Eq. (\ref{PsiQ}). We are interested in how the initial conditions endow $A$ with a time dependence.

\section{A Toy Model}\label{sec:delta}

As discussed in Section \ref{sec:intro}, we are interested in the Cauchy problem of Eq. (\ref{Vgeneral}) for which the relevant Green function is analytically known. This will allow us to explore the implications of a wide range of initial conditions, without having to contend with ambiguities arising from the fitting procedure for QNMs. We adopt as a toy model the delta function potential. In other words, we are interested in the solutions to the following equation:
\begin{eqnarray}
\label{Vgeneral}
(-\partial_t^2 + \partial_x^2 - V_0 \delta(x) ) \Psi(t, x) = 0 \, ,
\end{eqnarray}
where $V_0$ is a constant; $1/V_0$ sets the time (and spatial) scale of the problem.
One can think of the delta function as a very crude representation of
the Regge-Wheeler (or Zerilli) potential, with the peak of the potential located at $x=0$
($x$ is, up to an additive constant, the tortoise coordinate, with the horizon at $x=-\infty$).
\footnote{\label{xtortoise}
  Note that the peak location of the Regge-Wheeler potential is mildly $\ell$ dependent,
  but is at around $r = 3 M$, corresponding to a tortoise coordinate of $r_* \sim 1.6 M$. (Recall $r_* = r + 2M {\,\rm ln\,} [(r - 2M)/ 2M]$.)
  Thus, think of $x \sim r_* - 1.6 M$. 
}

Recall that this equation has the following formal solution:
\begin{align}
    \Psi (t,x) =\int d\bar{x} \left[ \partial_{\bar{t}}G
      |_{\bar{t}=0}\Psi_{0}(\bar{x})- G |_{\bar{t}=0}\dot{\Psi}_{0}(\bar{x})\right],\label{Phi1_Gral_B}
\end{align}
with $\Psi_0$ and $\dot \Psi_0$ being the perturbation and its time derivative
at the initial time $0$. Here, $G(t, x| \bar t, \bar x)$ is the retarded Green
function, which, for the delta function potential, can be straightforwardly
derived using standard methods
(see e.g. \cite{Hui:2019aox}):
\begin{align}
    &G(t,x|\bar{t},\bar{x}) =  G_F (t,x|\bar{t},\bar{x}) +  G_Q (t,x|\bar{t},\bar{x}),
\end{align}
where
\begin{align}
    & G_F (t,x|\bar{t},\bar{x}) = -\frac{1}{2}\left[\Theta(t-\bar{t}-|x-\bar{x}|)-\Theta( t  - \bar{t} - |x| - |\bar{x}|)\right], \label{GF_Delta}\\
    & G_Q (t,x|\bar{t},\bar{x}) = -\frac{1}{2}e^{-\frac{V_0}{2}(t-\bar{t}-|x|-|\bar{x}|)}  \Theta(t-\bar{t}-|x|-|\bar{x}|) \, .\label{GQ_Delta}
\end{align}
Here, $G_Q$ ($Q$ for quasinormal) represents the part of the Green function
that knows about the only linear QNM mode for the toy model,
with $\omega = -i V_0/2$ (i.e.\ the quasinormal frequency is purely imaginary).
$G_F$ represents the part of the Green function that does not depend
on $V_0$, and would be present even if the potential were zero. The subscript $F$
stands for ``flat space'', where the potential would be absent.

\begin{figure} [h!] 
    \centering
    {\includegraphics[width=0.43\textwidth]{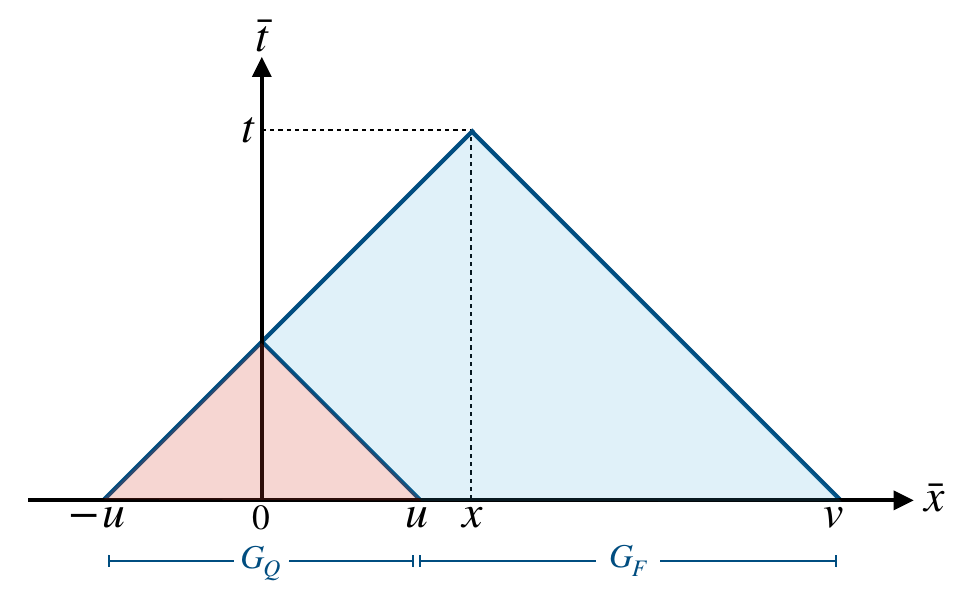}}
    \caption{
      Support of $G_F$ (shaded blue) and $G_Q$ (shaded red) for a point $(x, t)$. Here, $u = t-x$ and $v = t+x$. The potential peak is at $\bar x = 0$. The boundaries of these regions illustrate the role of causality constraints. Horizontal blue lines indicate the maximum size of the spatial region causally connected to $(x, t)$ through $G_Q$ and $G_F$, at the initial time $\bar t = 0$. Here, we have implicitly assume $t > x$
      (and $x > 0$). If $t < x$, the red region would be absent, which means no QNM is generated at $(x, t)$ starting
      from initial perturbations at $\bar t = 0$.} 
    \label{causality graph}
\end{figure}

The two step functions enforce two separate causality constraints.
The step function $\Theta(t-\bar{t}-|x-\bar{x}|)$ is non-zero only if
$(\bar x, \bar t)$ is within the past lightcone of $(x, t)$.
The step function $\Theta(t-\bar{t}-|x|-|\bar{x}|)$ is non-zero only if
$(\bar x, \bar t)$ can be causally connected to $(x, t)$ via a path that
touches/crosses the origin (location of the delta function). Thus, $G_Q$ is non-vanishing only in the red
region depicted in Figure \ref{causality graph}. This is the region from which
perturbations can scatter off the (delta function) potential before reaching the observer at $(x, t)$;
this is in fact how the QNM is created. The other part of the Green function $G_F$ is
non-vanishing only in the blue region depicted in Figure \ref{causality graph}.
This is the region from which perturbations propagate directly to $(x, t)$ without the possibility
of scattering off the (delta function) potential.
These causality constraints, suitably generalized, are present for more general potentials (see \cite{Szpak:2004sf}).

%

Given some initial conditions $\Psi_0(x)$ and $\dot{\Psi}_0(x)$, the total linear solution will contain two pieces, coming from $G_F$ and $G_Q$. In the former case, we replace Eq.\ (\ref{GF_Delta}) into Eq.\ (\ref{Phi1_Gral}) and obtain:
\begin{align}
    \Psi_F(t,x)&=\frac{1}{2}\left(\Psi_0(-u)+\Psi_0(v)\right)+\frac{1}{2}\int_{-u}^v d\bar{x} \;\dot{\Psi}_0(\bar{x})\nonumber\\
    &-\frac{1}{2}\Theta(t-|x|)\left(\Psi_0(|x|-t)+\Psi_0(t-|x|)\right)\nonumber\\
    & - \frac{1}{2}\Theta(t-|x|)\int_{|x|-t}^{t-|x|} d\bar{x} \;\dot{\Psi}_0(\bar{x}),\label{PsiF1_Delta}
\end{align}
where we have defined $u=t-x$ and $v=t+x$.

Next, we calculate the linear solution coming from $G_Q$.  Substituting Eq.\ (\ref{GQ_Delta}) into Eq.\ (\ref{Phi1_Gral}) we obtain:
\begin{align}
    \Psi_Q(t,x)&= A(t,x) e^{-\frac{V_0}{2}(t-|x|)}\Theta(t-|x|)\label{Psi1_GralAmplitude}\\
    & + \frac{1}{2}\left[\Psi_0(t-|x|) +\Psi_0(|x|-t) \right]\Theta(t-|x|)\,\label{Psi1_QFree} ,\\
    A(t,x) & =\frac{1}{2} \int_{|x|-t}^{t-|x|}d\bar{x}e^{\frac{V_0}{2}|\bar{x}|}\left[\dot{\Psi}_0(\bar{x})-\Psi_0(\bar{x})\frac{V_0}{2}\right].\label{A_Integral}
\end{align}
From here we see that the first line in $\Psi_Q$, given by Eq.\ (\ref{Psi1_GralAmplitude}), looks analogous to the usual QNM models used in the literature as it contains an exponential with the linear QNM frequency $\omega=-iV_0/2$ and the radiation is outgoing at spatial infinity and ingoing at the horizon. 

The total signal is then given by 
$\Psi=\Psi_F+\Psi_Q$ and one can see that line (\ref{Psi1_QFree}) cancels out the second line of Eq.\ (\ref{PsiF1_Delta}), which is necessary in order to recover free-space waves when $V_0=0$. Therefore, it will be convenient to redefine the total signal including only observable contributions as:
\begin{align}
    \Psi& = \tilde{\Psi}_F + \tilde{\Psi}_Q, \label{PsiTotal}\\
    \tilde{\Psi}_F&=\frac{1}{2}\left(\Psi_0(-u)+\Psi_0(v)\right)+\frac{1}{2}\int_{-u}^v d\bar{x} \;\dot{\Psi}_0(\bar{x})\nonumber\\
    & - \frac{1}{2}\Theta(t-|x|)\int_{|x|-t}^{t-|x|} d\bar{x} \;\dot{\Psi}_0(\bar{x}),\label{PsiF}\\
    \tilde{\Psi}_Q&= A(t,x) e^{-\frac{V_0}{2}(t-|x|)}\Theta(t-|x|)\label{PsiQ}.
\end{align}
Notice the step function in the quasi-normal part $\tilde{\Psi}_Q$: it tells us $\tilde{\Psi}_Q$ is non-zero only if the red-shaded region in Figure \ref{causality graph} exists. Indeed, the amplitude $A(t,x)$ 
(Eq. \ref{A_Integral})
is determined by an integral of the initial perturbation over the base of the red-shaded triangle (at the initial time).
In the following sections, we will focus on studying the behavior of $\tilde{\Psi}_Q$ for different choices of initial conditions. It's nonetheless worth keeping in mind that the
$\tilde{\Psi}_F$ contribution, from which the QNM is absent,
is generically non-vanishing.


\section{Elusive Quasinormal Modes} \label{sec:nQNMs}


In this section, we explore examples of initial conditions for which QNMs are elusive: either they fail to materialize, or if they do materialize, they would be difficult to identify and isolate. These examples might seem academic, since QNMs appear to be routinely identifiable from numerical simulations of binary mergers. In our view, the very fact that this is true tells us something interesting about the kind of initial conditions generated by binary mergers. We will have more to say about this below.

\subsection{No QNMs}
\label{No QNMs}

Firstly, as emphasized earlier, QNMs are generated only if the observer is causally connected to the peak of the potential barrier
at the initial time (i.e.\ merger time). 
We expect this condition to hold for all observationally relevant situations. 
Assuming so, note that the QNM amplitude $A(t,x)$ is determined by an integral over the initial condition, Eq. (\ref{A_Integral}). It is straightforward to see that if
the initial conditions were such that 
\begin{eqnarray}
\dot{\Psi}_0(\bar{x})-\Psi_0(\bar{x})\frac{V_0}{2}
\nonumber
\end{eqnarray}
is an odd function of $\bar x$ ($\bar x = 0$ being the location of the delta function potential), the QNM amplitude would vanish. 
This combination of $\Psi_0$ and $\dot{\Psi}_0$ at the initial time can in general be written as the sum of an even part and odd part. Provided that the even part is non-zero, one generically expects the QNM amplitude to be non-vanishing. Nonetheless, the purely odd example highlights the fact that even the existence of QNMs should not be taken for granted.

\subsection{Hidden QNMs}
\label{Hidden QNMs}

Here, we explore examples of initial conditions for which the amplitude $A(t,x)$ has such a strong time dependence that, the quasinormal time dependence $e^{-V_0 t/2}$ becomes hard to isolate in $\tilde{\Psi}_Q$ (Eq. \ref{PsiQ}). Recall again the integral for $A(t,x)$:
\begin{eqnarray}
A(t,x) & =\frac{1}{2} \int_{|x|-t}^{t-|x|}d\bar{x}e^{\frac{V_0}{2}|\bar{x}|}\left[\dot{\Psi}_0(\bar{x})-\Psi_0(\bar{x})\frac{V_0}{2}\right].
\label{A_Integral 2}
\end{eqnarray}
Generally, we expect the initial perturbation to vanish far away from the black hole, i.e.\ $\bar x \rightarrow \infty$, but it is possibly large close to the horizon $\bar x \rightarrow -\infty$.

As a first example, consider $\dot{\Psi}_0(\bar{x})-\Psi_0(\bar{x})\frac{V_0}{2}
= e^{-\alpha\bar x}$ with $\alpha > 0$. 
\footnote{Such an initial perturbation diverges at the horizon $\bar x \rightarrow -\infty$. Realistic initial conditions should not do that. We will have further comments on this below.
}
Performing the integration to obtain $A(t, x)$ and plugging into 
$\tilde{\Psi}_Q \propto A(t,x) e^{-V_0(t-|x|)/2}$
(c.f.\ Eq.\ (\ref{PsiQ})), we have,
\begin{equation}
\tilde{\Psi}_Q = \frac{-2 V_0 e^{\frac{1}{2} V_0 (| x| -t)}+(2 \alpha +V_0) e^{\alpha  (| x| -t)}+(V_0-2 \alpha ) e^{\alpha  (t-| x| )}}{V_0^2-4 \alpha ^2},
\end{equation}
where we have omitted the $\Theta$ function for concision. 

In this solution, the first term appears to be a true QNM, whereas the last two terms are not. In fact, given that $\alpha > 0$, it is possible for the third one to grow exponentially with time. 

Generally, as $t\rightarrow \infty$, the growing mode will dominate. Note, however, that depending on the relationship between $\alpha$ and $V_0$ (for example if $\alpha<V_0/2$), the magnitude of the QNM's coefficient can be larger than those of the two other terms, so the QNM mode may dominate at least initially (when $t\approx |x|$). Nevertheless, as $t\rightarrow \infty$, this QNM term decays and will be suppressed. So we can see the dominance of QNM solutions in the signal is not generally guaranteed, and in fact it is dependent on the properties of the initial conditions.

\begin{figure} [h!] 
    \centering
{\includegraphics[width=0.45\textwidth]{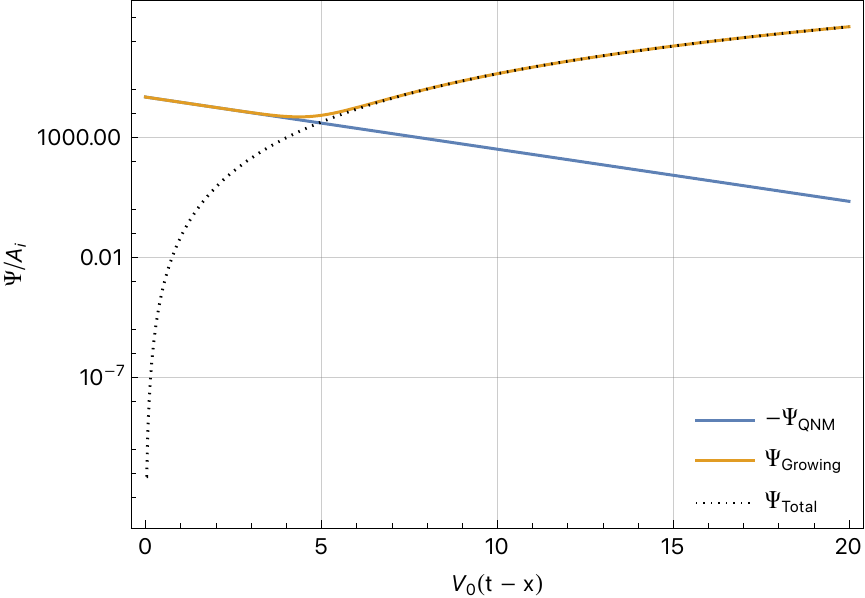}}
    \caption{Log plot comparing the QNM and growing contributions (\ref{eq:QNM3}) and (\ref{eq:Growing3}) against the total solution (\ref{totalgrw}) as a function of $V_0(t-x)$, for $n=3$. Here $x$ is the location of the observer and $t$ is the observation time. We are interested in cases where $x$ is large and positive (the observer is far from the black hole). For initial perturbation located around the delta function potential, the time by which the signal would arrive at the observer is around $t = x$. Hence, we plot the signal as a function of $V_0(t-x)$, using $1/V_0$ as the characteristic timescale (Table \ref{Params}).}  
    \label{piecewise}
\end{figure}

Let us consider another example: 
\begin{equation} \label{genPoly}
\dot \Psi_0(\bar x) - \Psi_0 (\bar x) {V_0 \over 2} =A_i
    \begin{cases}
        \bar x^{2n} & , -\infty < \bar x \leq 0\\
        0 & , 0 \leq \bar x < \infty 
    \end{cases}
,
\end{equation}
with an arbitrary overall scaling $A_i$.
Assuming that $n \geq 1$, we find the following solution,
\begin{align}
   \tilde{\Psi}_Q &= A_i\frac{4^{n}}{V_0^{2n+1}} e^{\frac{1}{2} V_0 (|x| - t)}\nonumber\\ & \times \left(-\Gamma[2n+1] + \Gamma [2n+1, \frac{1}{2} V_0 (|x| - t)]\right) ,
   \label{increasinGamma}
\end{align}
where, again, we have omitted the $\Theta$ function. Here, we introduced the incomplete gamma function $\Gamma[a, x] = \int_x^\infty s^{a-1} e^{-s} ds$. 
We split $\tilde {\Psi}_Q$ into two parts:
\begin{equation}
    \tilde{\Psi}_Q = \Psi_{\rm QNM} + \Psi_{\rm Growing},
    \label{totalgrw}
\end{equation}
where 
\begin{align} 
    \Psi_{\rm QNM} &\equiv -A_i\frac{4^{n}}{V_0^{2n+1}} e^{-\frac{V_0}{2}(t-|x|)} \Gamma[1+2n] ,
    \label{eq:QNM3} \\
    \Psi_{\rm Growing} &\equiv  A_i \frac{4^{n}}{V_0^{2n+1}} e^{-\frac{V_0}{2}(t-|x|)}  \Gamma[1+2n, -\frac{V_0}{2}(t-|x|)] .
    \label{eq:Growing3}
\end{align}
Here, $\Psi_{\rm QNM}$ has the expected QNM time-dependence; $\Psi_{\rm Growing}$, on the other hand, has a time-dependent (incomplete) gamma function that could potentially dominate over the QNM. As an illustration, in Figure \ref{piecewise}, we show the two contributions for $n = 3$. Initially, the contributions of the growing solution and the QNM mode are comparable. But after the early phase, which lengthens with increasing $n$, the growing solution dominates and the QNM becomes highly suppressed. 
As another example, if $n=0$, describing an initial perturbation that does not grow towards the horizon, the solution takes a simple form:
\begin{equation}
  \tilde{\Psi}_Q =  -\frac{1}{V_0} \left(e^{\frac{1}{2} V_0 (| x| -t)}-1\right),
  \label{PsiQn0}
\end{equation}
which corresponds to a QNM piece plus a constant, in agreement with  \cite{Lagos:2022otp}.


\vspace{0.3cm}

It's worth stressing that the examples considered in this section (\ref{sec:nQNMs}), while illustrating that a quasi-normal time dependence is far from guaranteed, are also largely unrealistic. 
We do not expect the initial perturbation to be purely odd, nor do we expect it to diverge at the horizon. Indeed, analyzes of numerical simulations of binary mergers routinely yield well-identified QNMs. This tells us the initial perturbation from such merger events (1) cannot take a special (odd) form that gives zero QNMs, and (2) must level off at some point, even as it tends to grow towards the horizon (see e.g. \cite{Bhagwat_2018,Okounkova:2020vwu}).
Keep in mind that we have been using $x$ as the proxy for the tortoise radius (up to an additive constant, see footnote \ref{xtortoise}), with $x \rightarrow -\infty$ representing the horizon. While simulations suggest the initial perturbation tends to be larger closer to the horizon, they should remain finite there. Thus, modeling the initial perturbation as $x^{2n}$ for $n \ge 1$ might be correct for some range of $x$, but eventually as $x \rightarrow -\infty$, we expect the effective $n$ to become zero. As a result, the behavior seen in Figure \ref{piecewise}, where $\Psi_{\rm Growing}$ hides the quasi-normal time-dependence, should be temporary and at sufficiently late times, we expect the quasi-normal time-dependence to emerge. We will see this expectation realized in the examples in the next two sections.

Let us summarize the main lesson so far: that the
spatial dependence of the initial perturbation has an impact on the time-dependence of the observed signal; in particular, an increase of the initial perturbation towards the horizon could lead to a (temporarily) growing amplitude for the QNMs.

\section{Gaussian Initial Condition}
\label{sec:gaussiansection}
Contrary to the usual QNM models assumed in the literature, the amplitude $A(t,x)$ in (\ref{A_Integral}) is not necessarily given by a constant, since the integration boundaries depend on $t$ and $x$ due to causality conditions. We emphasize that an evolving amplitude is not a sign of linear perturbation breaking down, but it instead provides information about the shape of the initial conditions around the potential peak. This amplitude evolution was discussed in \cite{Andersson:1996cm}, and also illustrated in a toy example in \cite{Szpak:2004sf}. Nonetheless, if the initial conditions are localized in a region smaller than $t-|x|$, then that region will determine the integration bounds and $A(t,x)$ will reach a constant for sufficiently large $t-|x|$. In what follows, we will illustrate these points with an example of a gaussian initial condition.

Let us consider an initial perturbation with compact support, given by a  Gaussian: 
\begin{equation}\label{gaussIC}
    \Psi_0(x)=A_i\exp\{ - (a (x-c))^2/4\}, \quad \dot{\Psi}_0=0,
\end{equation}
where we include a maximum amplitude $A_i$, a shift parameter $c$, and a decay width $a^{-1}$, in the initial condition model. Notice that, even though we will assume (\ref{gaussIC}), the same QNM solution $\tilde\Psi_Q$ (but not $\tilde \Psi_F$) would be obtained if $\dot{\Psi}_0\not=0$ and instead the combination $(\Psi_0-2\dot{\Psi}_0/V_0)$ took the Gaussian form considered in Eq.\ (\ref{gaussIC})
(see \ Eq.\ \ref{A_Integral}).
The quasinormal mode solution $\tilde{\Psi}_Q$ is found to be:
\begin{align} 
   & \tilde{\Psi}_Q = \theta (t-| x| ) \frac{\sqrt{\pi } V_0  A_i}{4 a} e^{-\frac{V_0}{2}(t-|x|+c)} e^{\frac{V_0^2}{4a^2}} \label{expPart}\\
    &\quad \times \left\{e^{c V_0} \left(\text{erfc}\left[\frac{a c}{2}+\frac{V_0}{2 a}\right]-\text{erfc}\left[\frac{a (| x| +c-t)}{2}+\frac{V_0}{2 a}\right]\right) \right. \label{2ndLine} \\
    &\quad \left. + \text{erfc}\left[\frac{a (-| x| +c+t)}{2}-\frac{V_0}{2 a}\right] + \text{erf}\left[\frac{a c}{2}-\frac{V_0}{2 a}\right]-1\right\},\label{QNM_gauss_first}
\end{align}
which is expressed in terms of the error and complementary error functions:
\begin{equation}
    \text{erf}(x) = \frac{2}{\sqrt{\pi}} \int_{0}^{x} e^{-t^2} \, dt; \; \text{erfc}(x) = \frac{2}{\sqrt{\pi}} \int_{x}^{\infty} e^{-t^2} \, dt = 1 - \text{erf}(x).\nonumber
\end{equation}
At asymptotically late times, when the argument of the \( \operatorname{erf}(x) \) function becomes large, it converges to a constant value of either -1 or 1, contingent on the sign of the argument. Similarly, for the \( \operatorname{erfc}(x) \) function, a large argument results in convergence to a constant value of either 0 or 2, again dependent on the sign of the argument. Throughout the text, we refer to this behavior as the functions attaining their respective constant asymptotic values. Clearly, the solution $\tilde{\Psi}_Q$ does not behave as a typical QNM for all times. Nonetheless,  we can take the late-time limit and obtain
\begin{align} 
    \label{QNM_gauss_first_latetime}
    \tilde{\Psi}_{Q,\text{late}} &= \theta (t-| x| ) \frac{\sqrt{\pi } V_0  A_i}{4 a} e^{-\frac{V_0}{2}(t-|x|+c)} e^{\frac{V_0^2}{4a^2}} \nonumber\\
    &\quad \times \left(e^{c V_0} \text{erfc}\left[\frac{a^2 c}{2a}+\frac{V_0}{2 a}\right] + \text{erf}\left[\frac{a^2 c}{2a}-\frac{V_0}{2 a}\right]+1\right), 
\end{align}
which does indeed behave as a typical QNM: it decays with the appropriate frequency ($V_0/2$) and has a constant amplitude. 
Note that this late-time limit reflects the fact that the error functions have reached their constant  asymptotic values. 

We emphasize that in Eqs.\ (\ref{expPart}) - (\ref{QNM_gauss_first}), the dependency on the ratio $V_0/a$ is important. It influences the constant prefactor, the arguments of the time-dependent terms, and the exponential $\exp\{V_0^2/(4a^2)\}$ in the first term. Next, we will investigate how this ratio influences the magnitude and dynamics of the QNM, particularly the duration until $\tilde{\Psi}_Q$ achieves a constant amplitude.

\begin{figure*}
\centering
\includegraphics[width=\textwidth]{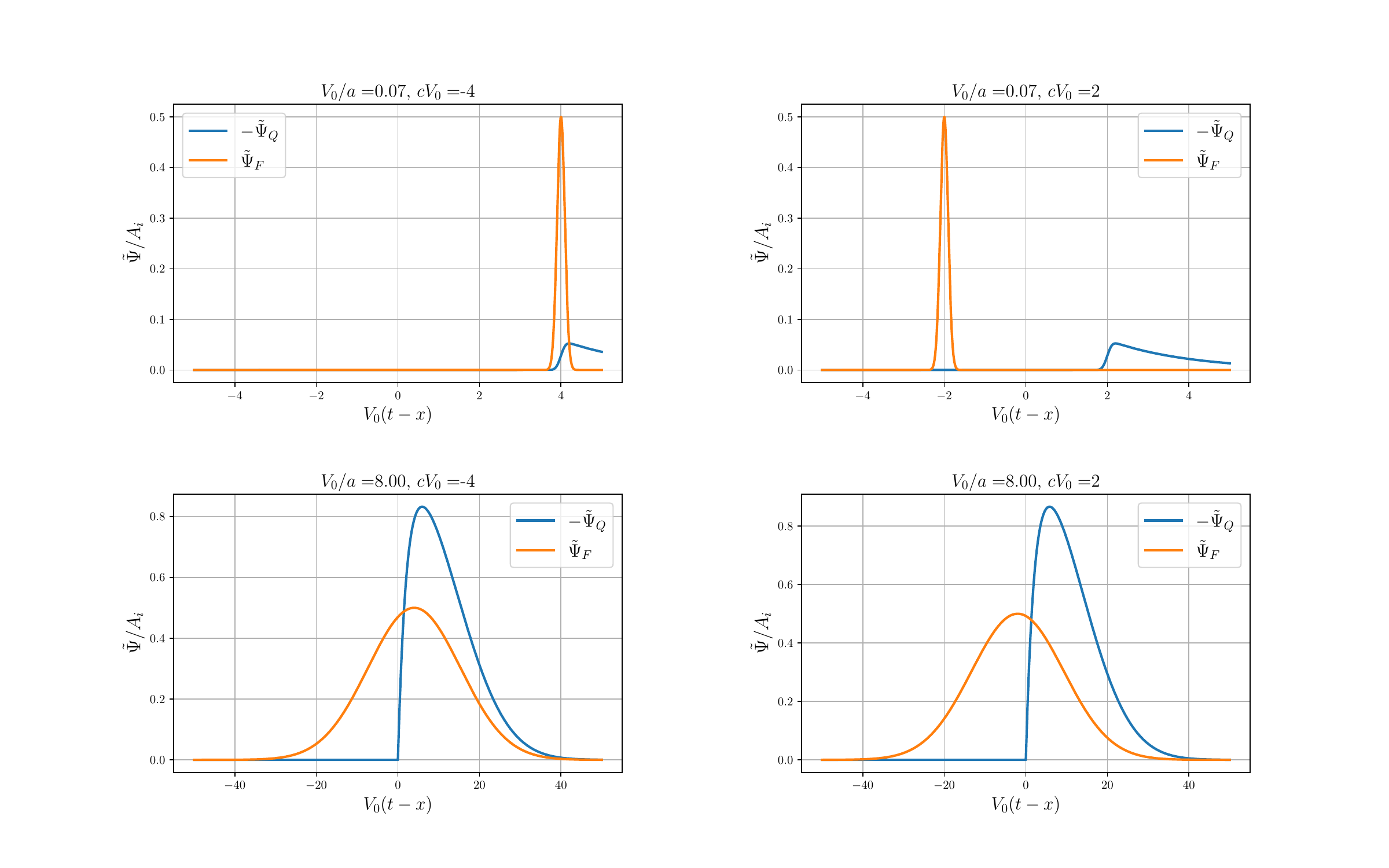}
    \caption{Comparison of $\tilde{\Psi}_Q$ (\ref{PsiQ}) and $\tilde{\Psi}_F$ (\ref{PsiF}) for a Gaussian initial condition. For plotting $\tilde{\Psi}_F$ we assumed it to be a pure function of $t-x$ valid for asymptotically far observers. See comments in the caption of Fig.\ \ref{piecewise} on why the signals are plotted as a function of $V_0 (t - x)$. \label{gaussian_comparison}    }  
\end{figure*}

\subsection{The Influence of Initial Condition}


In this section, we explore the properties of the solution (\ref{expPart})-(\ref{QNM_gauss_first}) and its dependence on the factor $V_0/a$. Physically, this would describe the width of the initial Gaussian in tortoise coordinates, in units of  $5.6 M$, and we find it to be the main quantity affecting the behavior of the QNM evolution. In Figure \ref{gaussian_comparison}, we illustrate with examples the expected signals for initial Gaussian perturbations of different widths: small $V_0/a$ means narrow, and large $V_0/a$ means wide.
%

The general intuition for the time dependence of the QNM amplitude is  that when the limits of integration
(Eq.\ \ref{A_Integral}), which vary with time, traverse the Gaussian initial perturbation, the amplitude of the QNM evolves in time. However, at sufficiently late times, when these limits encompass the majority of the Gaussian perturbation, the QNM amplitude asymptotically approaches a constant value. Consequently, a narrower Gaussian, characterized by a smaller $V_0/a$, results in a more rapid stabilization of the QNM amplitude. 
This intuition is substantiated by comparing the top two and bottom two graphs in Fig.\ \ref{gaussian_comparison}.


A few additional comments on the QNM signal $\tilde \Psi_Q$ are in order:
(1) It's clear there are terms in Eqs.\ (\ref{expPart}) to (\ref{QNM_gauss_first}) that correspond to a constant amplitude QNM, and terms that give rise to a time-dependent QNM amplitude (specifically, those involving the error function with time-dependent arguments). 
(2) The overall factor of $V_0/a$ out in front 
(Eq.\ \ref{expPart})
tells us that if the initial Gaussian perturbation has a narrow width (small $V_0/a$), the QNM amplitude is small. The suppression of the amplitude can be attributed to the integration over a narrow Gaussian profile (Eq.\ \ref{A_Integral}). (3) The appearance of $V_0/a$ in the exponent 
(Eq.\ \ref{expPart})
might seem to suggest an exponentially large amplitude for a large value of $V_0/a$ (a wide Gaussian), but that does not actually happen, due to cancellation from the combination of the error function terms.

We next turn our attention to the free part of the signal $\tilde \Psi_F$ (Eq.\ \ref{PsiF}). Recall that the total signal $\Psi$ is the sum of $\tilde \Psi_F$ and $\tilde \Psi_Q$
(Eq.\ \ref{PsiTotal}). 
While $\tilde \Psi_Q$ consists of signal that has undergone scattering with the potential (which is why its amplitude involves an integral of the initial perturbation around the location of the delta function potential), the free contribution $\tilde \Psi_F$ is roughly speaking the initial perturbation propagating directly to the observer. 
Indeed, assuming the initial perturbation vanishes far from the black hole (far right), and the observer at location $x$ is far from the black hole (and $\dot\Psi_0 = 0$) it can be shown from Eq.\ (\ref{PsiF}) that $\tilde \Psi_F = \Psi_0 (x-t)/2$, exactly the initial perturbation divided by $2$. This is what is shown in Fig.\ \ref{gaussian_comparison}. 
Based on the upper left panel of Fig.\ \ref{gaussian_comparison}, we see that, depending on the nature of the initial condition, there are time windows in which the free part dominates over the QNM part. In these cases, the behavior of the total observed signal will be mostly dependent on the initial condition.


Let us comment on the timing of the signal's arrival, and how it is related to the quantity $c V_0$. Recall that $c$ denotes the peak location of the initial perturbation, and $cV_0$ expresses the location 
in units of $5.6 M$. 
When the peak of the initial perturbation is positioned to the left of the potential (i.e.\ $cV_0<0$), both contributions 
$\tilde \Psi_F$ and $\tilde \Psi_Q$ arrive concurrently, for an observer at $x>0$ (as is always assumed).
Conversely, if the initial perturbation peaks to the right of the potential (i.e.\ $cV_0>0$), the quasinormal mode solution is expected to arrive later due to its causal connection to the potential (that is, the creation of $\tilde \Psi_Q$ requires scattering off the potential).
This is why, as can be seen in Fig.\ \ref{gaussian_comparison}, the arrival of $\tilde \Psi_F$ may precede that of $\tilde \Psi_Q$, but never vice versa. 
%
This is advantageous for the detection of quasinormal modes: by the time the QNM amplitude stabilizes, \(\tilde{\Psi}_F\) (which is independent of the QNM) will not be a significant contribution to the overall signal.




\subsection{Stabilization of the QNM amplitude}


In this section, we investigate in more detail the time it takes for the QNM amplitude to stabilize.

\begin{figure}[h!]
\centering
\includegraphics[width = 0.45\textwidth]{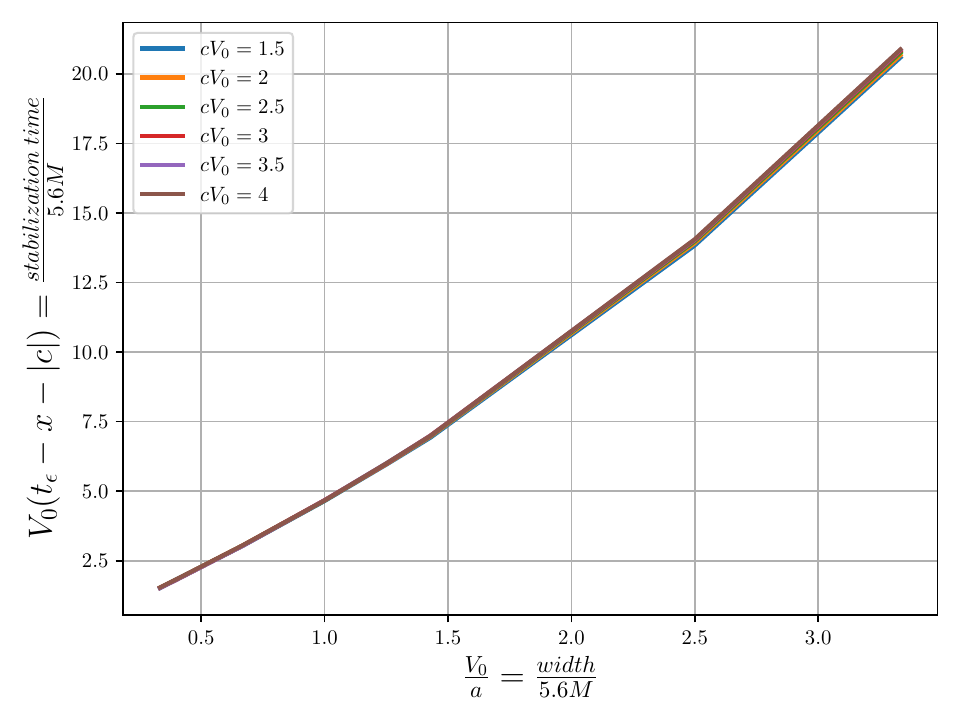}
 \caption{Graph of the time at which the QNM amplitude reaches a constant (stabilization time), for Gaussian initial perturbations. This is shown as a function of the Gaussian width
 $V_0/a$. 
 The Gaussian width determines the time it takes for the causality cone to encompass most of the initial perturbation, which is how we can intuitively understand the roughly linear relationship between the stabilization time and the initial perturbation width.
}
 \label{tthreshgaussian2}
\end{figure}

\begin{figure}[h!]
\centering
\includegraphics[width = 0.45\textwidth]{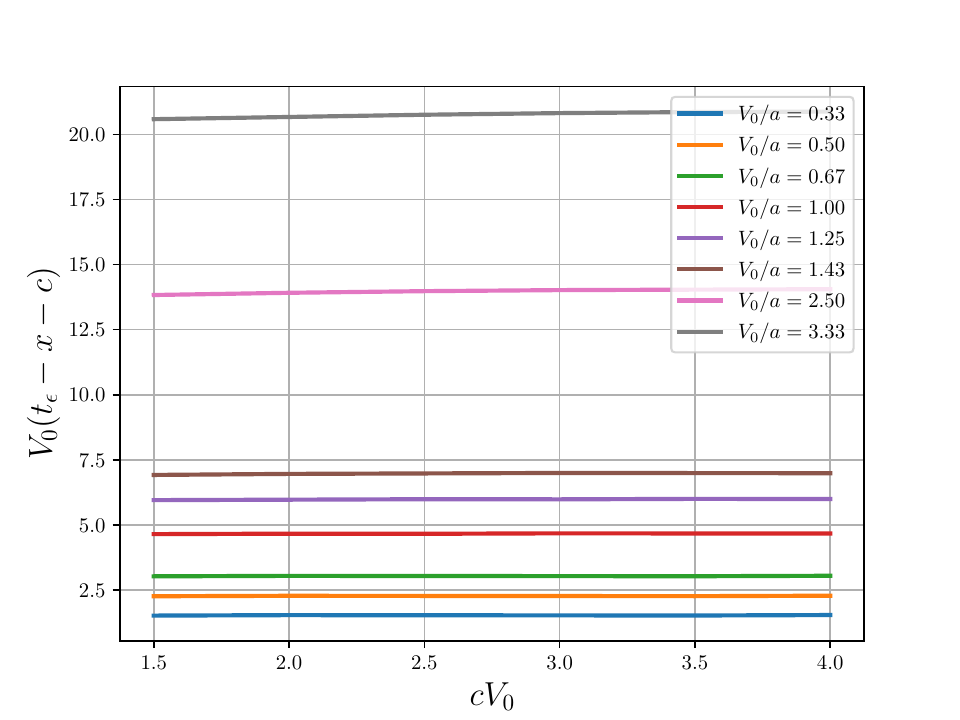}
 \caption{Graph of the time at which the QNM amplitude for initial Gaussian perturbations reaches a constant, varying the location and fixing the width of the Gaussian. 
 \label{tthreshgaussian1} }
\end{figure}

We introduce a small threshold, $\epsilon = \frac{A'}{A}$, to quantify amplitude stabilization for the quasinormal mode, with $A'$ being the time derivative of the amplitude and $A$ the amplitude itself. 
Practically, the amplitude is deemed stabilized when $\epsilon$ reaches $0.01$. The (dimensionless) time at which this is achieved is denoted by $V_0 t_\epsilon$. This is shown in
Figs.\ \ref{tthreshgaussian2} and \ref{tthreshgaussian1}, for different values of the width and central location of the Gaussian initial perturbation.



Figure \ref{tthreshgaussian2} shows $V_0(t_\epsilon-x-c)$ as a function of the width $V_0/a$, for positive values of $c$ (Gaussian peak location). The curve exhibits a nearly linear relationship between stabilization time and Gaussian width, which aligns with the previously stated intuition that broader initial conditions require more time for the quasinormal mode causal cone to fully encompass the initial condition. Furthermore, we see that the different color curves, representing different values of $c$, nearly overlap and thus have little effect, other than causing an overall delay in the signal's arrival time. 
Recall that $t=x+|c|$ is the time it takes for the QNM signal to travel from the peak of the Gaussian to an observer at $x>0$, keeping in mind that the QNM signal must first scatter off the peak of the potential (see Figure \ref{causality graph}).


  

Figure \ref{tthreshgaussian1} shows $V_0(t_\epsilon-x-c)$ as a function of $c V_0$ for different values of the 
Gaussian width $V_0/a$. It packages differently essentially the same information as in Figure \ref{tthreshgaussian2}.


Let us close with some remarks connecting our findings for the delta function toy model to observations of actual black holes. For concreteness, consider the $\ell=m=2$ fundamental mode ($n=0$) for a Schwarzschild black hole: its quasinormal frequency is
about $(0.37 - i \, 0.09) /M$. When fitting for QNMs, a common practice in the field is to start the fit at about $10 M$ after a binary merger (e.g. \cite{Buonanno:2006ui, Berti:2007fi}). This is comparable to the reciprocal of the imaginary part of the $(2,2,0)$ quasi-normal frequency, a not unreasonable approach if one is worried about significant nonlinearities close to the binary merger. Some recent studies, however, advocated fitting for QNMs right after merger \cite{Giesler:2019uxc, Isi:2019aib, Isi:2020tac}. 
The presence of nonlinearities close to merger would cast doubt on this approach
\cite{Buonanno:2006ui, Sberna:2021eui}, though it's conceivable that the nonlinearities (i.e.\ large perturbations) are confined to near-horizon and not near-light-ring regions most relevant for QNMs
\cite{Okounkova:2020vwu,Lagos:2022otp}.

What lessons can we draw from our toy model computations concerning this whole discussion? Equating $-V_0/2$ with the imaginary part of the $(2,2,0)$ quasi-normal frequency, we can think of $1/V_0$ as representing $5.6 M$ (Table \ref{Params}). From Figs.\ \ref{tthreshgaussian2} and \ref{tthreshgaussian1}, we see that the stabilization time would be about $10 M$ (i.e.\ $V_0 (t_\epsilon - x - c) \sim 1.8$),
if the tortoise width of the initial Gaussian perturbation were about $2.2 M$ (i.e.\ $V_0/a \sim 0.4$). 
Importantly, the stabilization time could be much longer if the initial Gaussian perturbation were wider. For instance, a tortoise width of 
$5.6 M$ ($V_0/a \sim 1$) would imply a stabilization time around $26 M$ ($V_0 (t_\epsilon - x - c) \sim 4.6$); a tortoise width of 
$14 M$ ($V_0/a \sim 2.5$) would imply a stabilization time around $78 M$ ($V_0 (t_\epsilon - x - c) \sim 14$). 
\footnote{In this discussion, we tacitly translate $1/V_0$ to $5.6 M$ by focusing on the dominant, $(2,2,0)$, mode. One could instead do the translation for other modes (of a higher angular momentum or overtone), which would in general shorten the timescale. Keep in mind though that the width of the relevant initial perturbation for these other modes might well be different from that for $(2,2,0)$. 
}

{\it To conclude:} (1) Even if nonlinearities were irrelevant (i.e.\ linear perturbation theory applies), it takes time for the QNM amplitude to stabilize. (2) The stabilization time increases as the initial perturbation width increases. The rule-of-thumb wait time of about $10 M$ in QNM fitting is by no means guaranteed to be the appropriate choice. The choice of $10 M$ is often justified by verifying that the resulting QNM amplitude does not change when the wait time is increased. While this seems a reasonable procedure, it would be reassuring to know from black hole merger simulations that the width of the initial perturbation indeed justifies the rule-of-thumb wait time. This would be helpful, especially in light of the possibility that different fitting procedures could lead to different conclusions \cite{Giesler:2019uxc,Baibhav:2023clw}).

\section{S-shaped Initial Condition}\label{sec:stepIC}


\begin{figure}[h!]
    \centering
    {\includegraphics[width=0.4\textwidth]{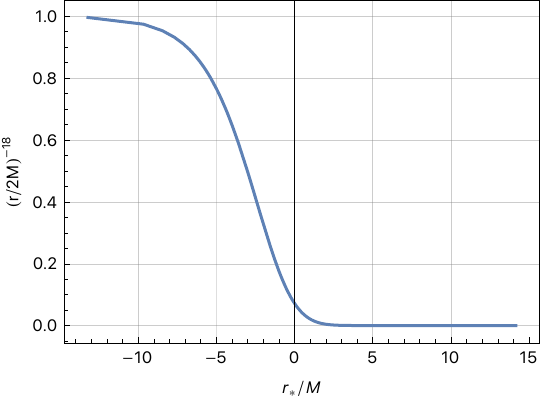}}
    \caption{Function $r^{-18}$ expressed in tortoise coordinates $r_*$. The light ring is located at $r_*\approx 1.6 M$.} 
    \label{powerlawdecay}
\end{figure}

\begin{figure}[h!]
    \centering
 {\includegraphics[width=0.45\textwidth]{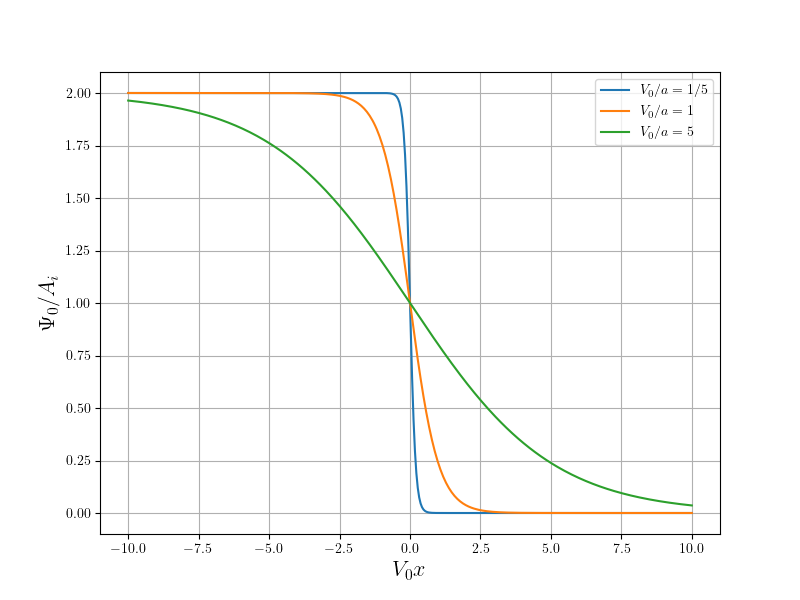}}
    \caption{Shape of the tanh initial condition, setting $c=0$ for various widths, with $V_0/a$ representing the dimensionless width, and $cV_0$ determining the dimensionless location of the ``step". Recall that $x$ can be interpreted roughly as $r_* \approx 1.6 M$, and $V_0 x$ represents the dimensionless spatial coordinate. 
    \label{tanhGraph} }    
\end{figure}

While the Gaussian initial condition studied in the previous section was instructive in illustrating how the observed signal depends on details of the initial perturbation, in this section, we wish to study a model that more closely resembles the aftermath of a binary merger, at least in one respect: that is the perturbation appears to be largest at the horizon. 
From numerical merger simulations, the initial perturbation tends to be enhanced close to the remnant black hole's horizon, going roughly as $r^{-n}$ with a rather large $n$ 
\cite{Bhagwat_2018}. 
As an example, consider an initial perturbation that goes as $r^{-18}$ in Schwarzschild radial coordinate (as estimated for one of the Kerness measures studied in \cite{Bhagwat_2018}), in tortoise coordinate it would take on an S-shaped profile, as depicted in Fig.\ \ref{powerlawdecay}. This motivates us to consider the following initial condition:
\begin{equation}
    \Psi_0(x)= A_i [1-\tanh(a (x-c))], \quad \dot{\Psi_0}=0,
\end{equation}
which asymptotes to a non-zero value $A_i$ towards the horizon, and vanishes far away from it, see Fig.\ \ref{tanhGraph}. 
The parameters $a$ and $c$ determine the width and location of the transition to zero. 
We again emphasize that the predicted $\tilde \Psi_Q$ remains identical even if $\dot\Psi \ne 0$, as long as the combination $\Psi_0 - 2\dot\Psi_0 / V_0$ takes the same form as above, but the predicted $\tilde\Psi_F$ would differ.

The analytic solution for $\tilde\Psi_Q$ is:
\begin{align} \label{PsiQtanh}
    \tilde{\Psi}_Q &=  A_i \theta (t-| x| ) \left( {}_2F_1\left[1,\frac{V_0}{4 a};\frac{V_0}{4 a}+1;-e^{-2 a c}\right] \right. \nonumber \\
    &\quad \left. +  {}_2F_1\left[1,-\frac{V_0}{4 a};1-\frac{V_0}{4 a};-e^{-2 a c}\right]\right) e^{\frac{1}{2} V_0 (| x| -t)} \nonumber \\
    &\quad \left. -  {}_2F_1\left[1,\frac{V_0}{4 a};\frac{V_0}{4 a}+1;-e^{-2 a (c-t+| x| )}\right] \right. \nonumber \\
    &\quad \left. -  {}_2F_1\left[1,-\frac{V_0}{4 a};1-\frac{V_0}{4 a};-e^{-2 a (c+t-| x| )}\right] \right),
\end{align}
where ${}_2F_1 (a,b;c;z)$ denotes the hypergeometric function. The first two hypergeometric functions describe a conventional constant-amplitude QNM contribution, while the latter two introduce a more complex time-dependence.

\begin{figure*}
    \centering  {\includegraphics[width=\textwidth]{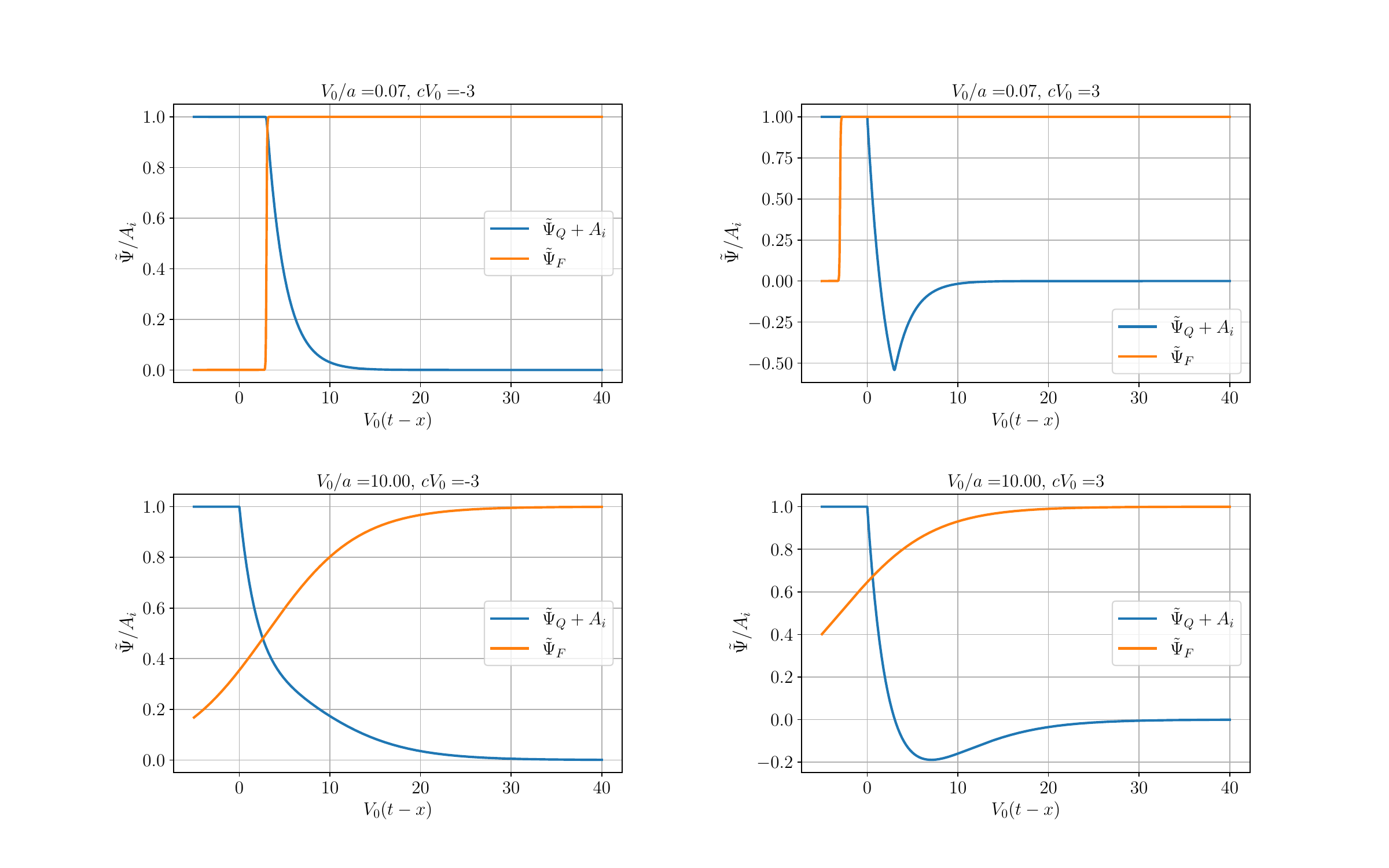}}
    \caption{Comparison between $\tilde{\Psi}_Q$ and $\tilde{\Psi}_F$.  $\tilde{\Psi}_F$ is essentially the unfiltered initial perturbation, i.e.\ $\tilde{\Psi}_F = \Psi_0 (x-t)/2$. 
    For the tanh initial condition at hand, the perturbation is large near the horizon and diminishes far from the black hole.
    See text for comments on $\tilde \Psi_Q$. \label{tanh comparisons} } 
\end{figure*}

In particular, let us study the late-time behavior of this solution. When $t\rightarrow \infty$, the last argument of the hypergeometric functions, $\exp\{-2a(c\pm t\mp |x|)\}$,  either goes to 0 or infinity depending on the sign of the argument of the exponential. If it is the former, the hypergometric function approaches one, but if it is the latter, then the function tends towards a combination of gamma functions that simplifies in a relatively nice way. For a generic constant $f$, as $z\rightarrow \infty$,

\begin{align}
    _2F_1(1,f;f+1;-z)&\rightarrow  z^{-f} \Gamma (1-f) \Gamma (f+1)\nonumber \\ 
    &+\frac{\Gamma (f-1) \Gamma (f+1)}{z \Gamma (f)^2}\nonumber \\ 
    &=\pi  f z^{-f} \csc [\pi  f]+\frac{f}{(f-1) z},
\end{align}
which implies that the late-time behavior for (\ref{PsiQtanh}) is the following:
\begin{align} 
    \tilde{\Psi}_{Q, \text{late}} &= A_i \theta (t-| x| ) \left( -1 - \frac{V_0 e^{2 a (| x| +c-t)}}{V_0-4 a} \right. \nonumber \\
    &\quad - \frac{\pi V_0 \csc \left(\frac{\pi V_0}{4 a}\right) e^{\frac{1}{2} V_0 (| x| +c-t)}}{4 a} \nonumber \\
    &\quad +\left( {}_2F_1\left[1,\frac{V_0}{4 a};\frac{V_0}{4 a}+1;-e^{-2 a c}\right] \right. \nonumber \\
    &\quad \left. \left. +\, {}_2F_1\left[1,-\frac{V_0}{4 a};1-\frac{V_0}{4 a};-e^{-2 a c}\right] \right)e^{\frac{1}{2} V_0 (| x| -t)} \right).\label{lattimeTanh}
\end{align}
This form for the late-time asymptotic behavior is valid as long as
$V_0/4a$ is not an integer (such that the cosecant remains finite). This is sufficient for our purpose, since there's no reason to expect $V_0/4a$ to be finely tuned to an integer value. 

In this late-time expression, (\ref{lattimeTanh}), there are two kinds of time dependent exponential terms: one that corresponds to the correct QNM frequency 
$e^{-V_0 t/2}$ and one that decays as 
$e^{-2a t}$, so the total solution does not uniformly behave as a traditional QNM. 
The $e^{-2a t}$ behavior reflects the shape of the initial perturbation, rather than the shape of the potential (a delta function controlled by $V_0$ alone).

Note also how $\tilde\Psi_{Q,{\rm late}}$ asymptotes to a constant ($-A_i$) at late times, as opposed to zero in the case of the Gaussian initial perturbation. This can be traced to the fact that the initial perturbation does not vanish at the horizon, unlike in the Gaussian case. This is reminiscent of gravitational memory, in which the gravitational wave signal does not return to zero after the wave is long gone. 



In Fig.\ \ref{tanh comparisons}, we show
$\tilde{\Psi}_Q$ and $\tilde{\Psi}_F$ for different choices of the (dimensionless) width $V_0/a$ and location $cV_0$ of the S-shaped feature represented by the hyperbolic tangent. It's helpful to remember that $\tilde \Psi_F$ is related to the initial perturbation by $\tilde \Psi_F = \Psi_0(x-t)/2$, assuming the observer at $x$ is located far from the black hole. 
Note also we plot $\tilde \Psi_Q + A_i$, which asymptotes to zero at late times.

The upper panels of Fig.\ \ref{tanh comparisons} show examples for a small value of $V_0/a$, for which the step in the hyperbolic tangent is relatively narrow. In this case, the $e^{-2at}$ contribution to $\tilde \Psi_{Q,{\rm late}}$ can be ignored, and one does find the expected QNM time dependence $e^{-V_0 t/2}$ at late times. The difference between the left and right panels has to do with whether the step is to the left or right of the delta function potential (effectively the location of the light ring). Notice how, if the step is to the right (positive $cV_0$), the initial perturbation is non-zero at the delta function potential and so $\tilde \Psi_Q$ starts deviating from zero as soon as $V_0 (t-x) > 0$. If the step is to the left (negative $cV_0$), the initial perturbation is essentially zero at the location of the delta function potential, and one has to wait for the step to come within view of the observer before $\tilde \Psi_Q$ switches on from zero.

The lower panels illustrate the behavior for a large value of $V_0/a$, for which the step in the hyperbolic tangent is relatively wide and gradual. In this case, $\tilde \Psi_Q$ takes longer to stabilize (i.e.\ to reach the late time asymptote $\Psi_{Q, {\rm late}}$). Moreover, the late time dependence is no longer dominated by the expected QNM $e^{-V_0 t /2}$, but by $e^{-2at}$ which reflects the (wide) S-shape initial perturbation. There is less of a difference between positive and negative $cV_0$, because the step is sufficiently wide that the initial perturbation is non-vanishing at the location of the delta function regardless. 

The dividing line between domination by $e^{-V_0 t /2}$ versus $e^{-2at}$ at late times is whether $V_0/a$ is smaller or larger than $4$. Recalling our heuristic of regarding $1/V_0$ as roughly equivalent to $5.6 M$. Thus, domination by the QNM requires the (tortoise) width $1/a$ of the initial perturbation be smaller than $22.4 M$. It would be useful to check whether this holds in simulations of black hole mergers.

\begin{figure}[h!]
\centering
\includegraphics[width = 0.45\textwidth]{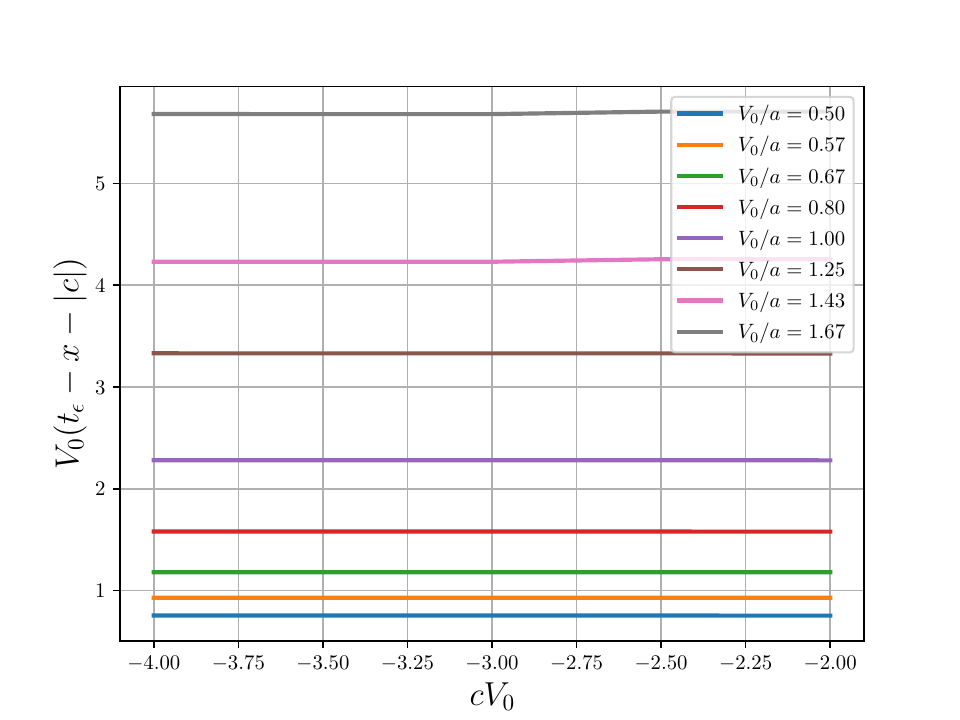}
 \caption{Stabilization time as a function of (negative) $c V_0$, for various values of $V_0/a$.
 \label{tthreshtanh1_minus} } 
\end{figure}

\begin{figure}[h!]
\centering
\includegraphics[width = 0.45\textwidth]{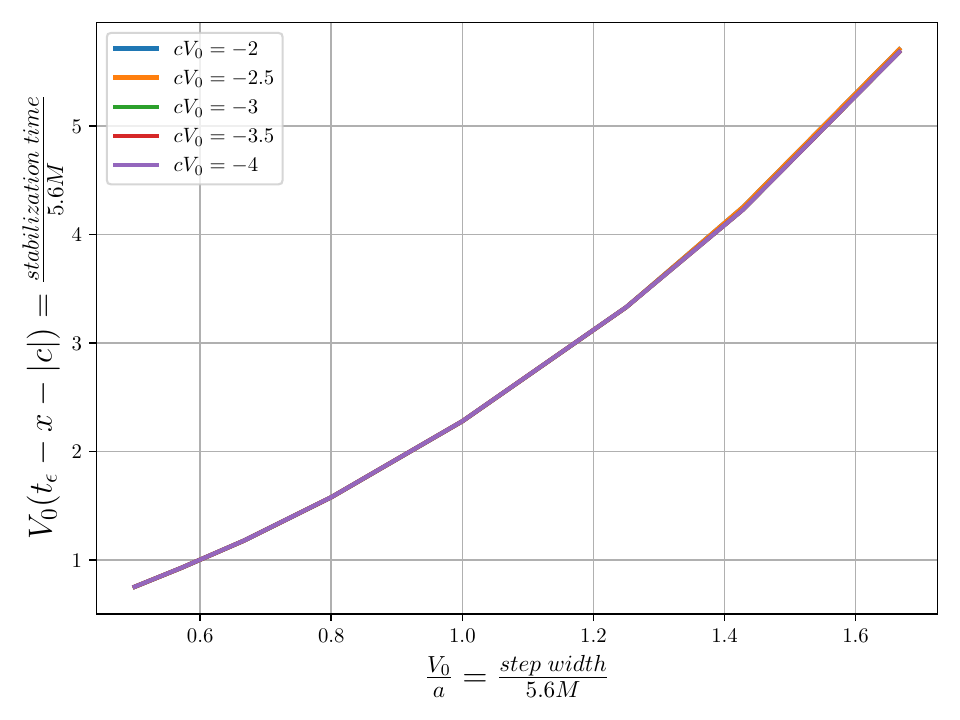}
 \caption{Stabilization time as a function of $V_0/a$, for various values of (negative) $c V_0$.
 } 
 \label{tthreshtanh2_minus}
\end{figure}

In Figs.\ \ref{tthreshtanh1_minus} and \ref{tthreshtanh2_minus}, we show the stabilization time as a function of the step location and width of the hyperbolic tangent initial perturbation. Previously, for a Gaussian initial perturbation, we define the stabilization time as the time it takes for the QNM amplitude to asymptote to a constant. For the hyperbolic tangent initial condition at hand, since the late time behavior is no longer guaranteed to go as $e^{-V_0 t/2}$, we instead define the stabilization time by asking when
$\epsilon \equiv (\tilde \Psi_{Q, {\rm late}} - \tilde \Psi_Q)/(\tilde \Psi_Q + A_i)$ drops below $1\%$. In other words, this is the time by which $\tilde \Psi_{Q, {\rm late}}$ is a good approximation for $\tilde \Psi_Q$. 


As expected, the stabilization time increases for larger widths $V_0/a$ of the initial perturbation and is relatively insensitive to the location $cV_0$ of the step. 
For the stabilization time to be about $10 M$ (i.e.\ $V_0 (t_\epsilon - x - |c|) \sim 1.8$), the tortoise width $1/a$ should be about $4.8 M$
(i.e.\ $V_0/a \sim 0.86 $). 
Figs.\ \ref{tthreshtanh1_minus} and \ref{tthreshtanh2_minus} are for negative values of $cV_0$, consistent with expectation based on Fig.\ \ref{powerlawdecay}.

\begin{figure}[h!]
\centering
\includegraphics[width = 0.45\textwidth]{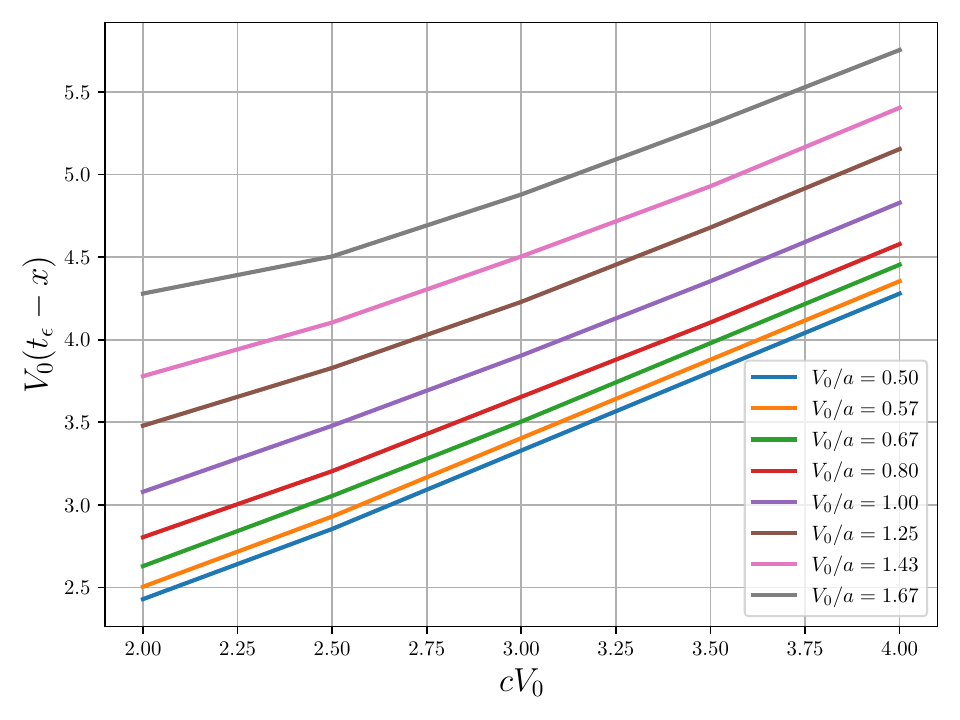}
 \caption{Analog of Fig.\ \ref{tthreshtanh1_minus} for positive $c V_0$.}
 \label{tthreshtanh1}
\end{figure}

\begin{figure}[h!]
\centering
\includegraphics[width = 0.45\textwidth]{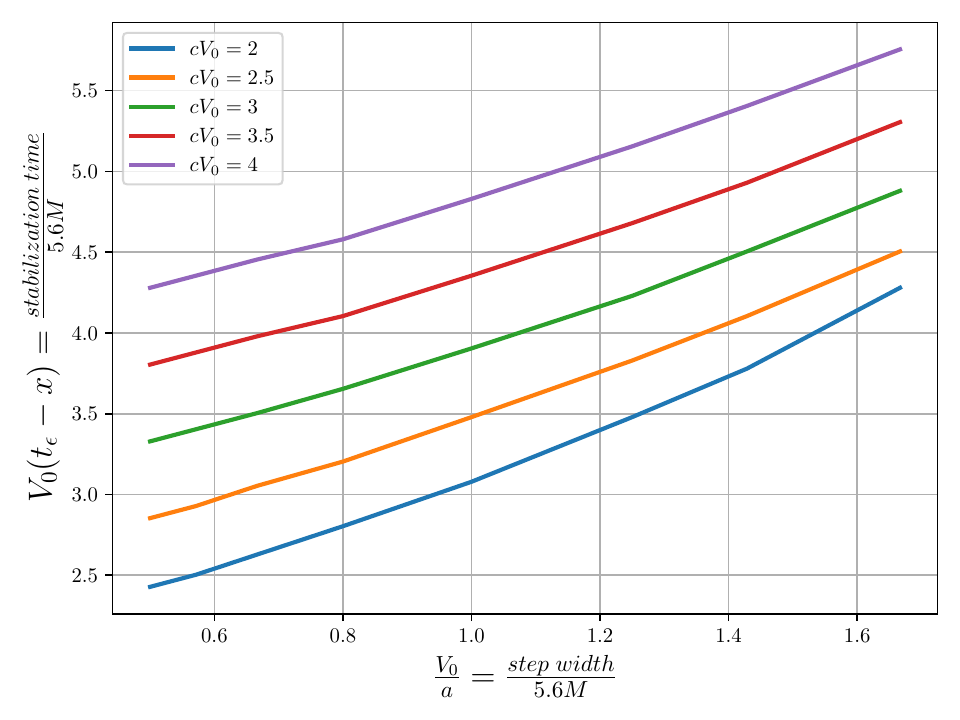}
 \caption{Analog of Fig.\ \ref{tthreshtanh2_minus} for positive $c V_0$. 
 } 
 \label{tthreshtanh2}
\end{figure}

For completeness, Figs.\ \ref{tthreshtanh1} and \ref{tthreshtanh2} show the stabilization time for positive values of $cV_0$. In this case, we quote $V_0(t-x)$, instead of $V_0(t-x-|c|)$, in the vertical axis because observers will see a non-zero signal starting immediately from $V_0(t-x)=0$, unlike the case of $V_0c<0$. From Fig.\ \ref{tthreshtanh1} we see that the stabilization time grows linearly with the value of $cV_0$, because the observer has to wait about $V_0(x+c)$ for the tanh transition to reach them and hence reach stability. From Fig.\ \ref{tthreshtanh2} we see that the stabilization time grows with the step width, as expected. 



Finally, it is worth noting that we have studied other s-shaped initial conditions such as the logistic, gudermannian, and inverse tangent functions, which give rise to similar behaviors as the hyperbolic tangent.

\section{Conclusions}\label{sec:conclusions}

In this paper, we have analyzed the impact of initial conditions on linear black hole ringdown gravitational wave signals. We have done this by using
a toy model for which the analytical solutions are known, and the quasi-normal mode can be unambiguously identified. This helps us sidestep difficulties in interpreting results from fitting QNMs to 
numerical merger simulations, where conclusions about the QNMs could be sensitive to details of the fitting procedure.

In this toy model, the effective potential generated by a black hole is replaced by a delta function.
The general ringdown signal has two physical contributions: the quasi-normal solution $\tilde\Psi_Q$ which comes from interactions of the initial perturbations with the potential, and the free solution $\tilde\Psi_F$ which comes from free traveling waves of the initial perturbations (see Eqs.\ (\ref{PsiTotal})-(\ref{PsiQ})). A key aspect we focus on is the fact that the quasi-normal solution $\tilde\Psi_Q$ comes with a {\it time-dependent} amplitude in general (after factoring out the exponential with a complex frequency).
We emphasize this is true even within the regime of validity of linear perturbation theory (see Figs.\ \ref{QNMspacetime} and \ref{causality graph} for an intuitive understanding of how this comes about). Contrast this with 
the standard QNM fitting of ringdown data, which typically assumes each QNM comes with a {\it constant} amplitude as long as linear perturbation theory applies.

We started our investigation by showing in Sec.\ \ref{sec:nQNMs} that initial conditions can be found which yield no discernible QNMs. Such initial conditions are admittedly ad hoc, but they underscore the fact that the properties of QNMs (including even their existence) are sensitive to the nature of the initial perturbations.
%
%

We then proceeded to consider two specific classes of initial perturbations, and their corresponding ringdown signals: Gaussian initial condition in Sec.\ \ref{sec:gaussiansection}, and S-shaped initial condition in Sec.\ \ref{sec:stepIC}. 
We found that depending on the parameters of the initial conditions, it is possible for $\tilde\Psi_F$ to dominate over $\tilde\Psi_Q$, at least for a certain period of time.
Furthermore, we showed that while these two classes of initial conditions do lead eventually to QNM with a constant amplitude at late times, the time it takes for $\tilde\Psi_Q$ to reach such an asymptote (i.e. the stabilization time) depends greatly on the width of the initial perturbation. In the cases we explored, the stabilization time ranged from $1 M$ to $110 M$: the larger the initial perturbation width, the longer the stabilization time. This result is relevant for  the standard QNM fitting of ringdown data (with QNMs of constant amplitudes); such fitting should only be done after the stabilization time in order to avoid biases. We emphasize again that this is true even within the regime of validity of linear perturbation theory. 


Let us close by briefly commenting on the implication of our findings for higher order perturbations. Consider for instance an equation of the form:
\begin{equation}
    (-\partial_t^2 + \partial_x^2 - V_0 \delta(x))\Psi = (\partial_t \Psi)^2 \, .
\end{equation}
We expect an equation of this schematic form (with a different potential of course) by expanding the Einstein equation to second order in perturbations around a BH. Expressing
$\Psi = \Psi_{(1)} + \Psi_{(2)}$, with $\Psi_{(2)}$ of order $\Psi_{(1)} {}^2$, we see that the linear perturbation $\Psi_{(1)}$ satisfies Eq.\ (\ref{Vgeneral}), and the quadratic perturbation $\Psi_{(2)}$ satisfies:
\begin{equation}
\label{Psi21}
    (-\partial_t^2 + \partial_x^2 - V_0 \delta(x))\Psi_{(2)} = (\partial_t \Psi_{(1)})^2 \, .
\end{equation}
This equation can be solved using the Green's function $G$ as (see e.g. \cite{Lagos:2022otp}):
\begin{equation}
\Psi_{(2)} (t, x) = \int d \bar t d\bar x \,\, G(t, x | \bar t , \bar x) \,
(\partial_{\bar t} \Psi_{(1)} (\bar t, \bar x) )^2 \, .
\end{equation}
The point is that no matter how long one waits, $\Psi_{(2)}$ observed at large $t$ is always influenced to some extent by $(\partial_{\bar t} \Psi_{(1)} (\bar t, \bar x) )^2$ at small $\bar t$. At such an early $\bar t$, there is no reason to expect $\Psi_{(1)}$ to consist purely of a linear QNM of a {\it stabilized} amplitude. 
Recent computations of quadratic QNMs
\cite{Bourg:2024jme, Bucciotti:2024zyp,Bucciotti:2024jrv,Ma:2024qcv} make just such an assumption: that, effectively, the right-hand side of Eq.\ (\ref{Psi21}) consists of the product of stabilized linear QNMs at all times. It would be useful to relax this assumption, and study whether doing so might close the gap between results from such computations and from numerical simulations. We hope to pursue this in the future. 



\section{ACKNOWLEDGMENTS}
A.C.\ is supported by the National
Science Foundation Graduate Research Fellowship under Grant No. DGE-2036197. M.L.\ was funded by the Innovative Theoretical Cosmology Fellowship, at Columbia University, for part of this research. L.H.\ acknowledges support from the DOE DE-SC011941 and a Simons Fellowship in Theoretical Physics.

\appendix

\bibliographystyle{apsrev4-1}
\bibliography{References.bib}

\end{document}